\documentclass[aps,pra,twocolumn,superscriptaddress]{revtex4}%
\usepackage{amsfonts}
\usepackage{amsmath}
\usepackage{amssymb}
\usepackage{graphicx}
\usepackage{float}
\newcommand{\qed}{\nobreak \ifvmode \relax \else
\ifdim\lastskip<1.5em \hskip-\lastskip
\hskip1.5em plus0em minus0.5em \fi \nobreak
\vrule height0.75em width0.5em depth0.25em\fi}

\begin{document}
\title{Quantum key distribution with phase-encoded coherent states: \\Asymptotic security analysis in thermal-loss channels}
\author{Panagiotis Papanastasiou}
\affiliation{Computer Science and York Centre for Quantum Technologies, University of York,
York YO10 5GH, United Kingdom}
\author{Cosmo Lupo}
\affiliation{Computer Science and York Centre for Quantum Technologies, University of York,
York YO10 5GH, United Kingdom}
\author{Christian Weedbrook}
\affiliation{Xanadu, 372 Richmond St W, Toronto, M5V 2L7, Canada}
\author{Stefano Pirandola}
\affiliation{Computer Science and York Centre for Quantum Technologies, University of York,
York YO10 5GH, United Kingdom}

\begin{abstract}
We consider discrete-alphabet encoding schemes for coherent-state
quantum key distribution. The sender encodes the letters of a
finite-size alphabet into coherent states whose amplitudes are
symmetrically distributed on a circle centered in the origin of
the phase space. We study the asymptotic performance of this
phase-encoded coherent-state protocol in direct and reverse
reconciliation assuming both loss and thermal noise in the
communication channel. In particular, we show that using just four
phase-shifted coherent states is sufficient for generating secret
key rates of the order of $4 \times 10^{-3}$ bits per channel use
at about $15$~dB loss in the presence of realistic excess noise.
\end{abstract}

\maketitle

\section{Introduction}
Quantum cryptography, or more accurately known as quantum key
distribution (QKD), is based on the laws of quantum
information~\cite{NiCh,Hayashi} to provide in principle secure
communication between two authorized
parties~\cite{Gisin2002,Scarani2008}, traditionally called Alice
and Bob. In particular these two parties exchange many signals
using a quantum channel which prohibits an exact duplication of
them~\cite{no cloning}. This fact allows the remote parties to
quantify and bound the amount of information that a potential
eavesdropper (Eve) may intercept, so that they can still extract
and share a secret key. Such key may then be used for data
encryption by means of the one-time pad~\cite{schneider}.

Since the first QKD protocol~\cite{BB84}, many advances have been
made including theoretical proofs, proof-of principle experiments
and in-field tests. Despite these efforts, the performance of any
point-to-point QKD protocol cannot surpass the fundamental
repeater-less PLOB bound established in Ref.~\cite{PLOB15} based
on the relative entropy of entanglement of the channel (see
Ref.~\cite{strong3} for a review, and Ref.~\cite{net} for an
extension to repeaters and arbitrary networks). However, it is
also true that continuous-variable (CV) QKD~\cite{Weedbrook2012}
has a key rate performance which is not far from this ultimate
bound when we assume ideal reconciliation and detectors with high
efficiency. Furthermore, another advantage of CV
systems~\cite{Braunstein} relies on the use of cheap room
temperature equipment, easily integrable in the current
telecommunication infrastructure.

In recent years, we have witnessed the introduction of many
protocols based on a CV encoding, e.g., exploiting a Gaussian
modulation of the amplitude of Gaussian states. These protocols
were designed for squeezed states~\cite{Nicolas,Hillery}, coherent
states~\cite{GG02,weedbrook2004noswitching}, thermal
states~\cite{filip-th1,weed1,usenkoTH1,weed2}, and also extended
from one-way to two-way quantum
communication~\cite{weed2way,pirs2way,2way2modes,QIprot1,flood1}
or reduced to one-dimensional encoding~\cite{1dim}. In addition,
protocols such as in
Ref.~\cite{CV-MDI-QKD,CVMDIQKD-reply,Ottaviani2015}, assuming
measurement-device-independence (MDI)~\cite{side-channel
attacks,MDILo} as a counter-measure against detectors'
side-channel attacks, have extended the concept of CV-QKD to
end-to-end network implementations~\cite{starprotocol}. For most
of these protocols, not only experiments were
shown~\cite{Grosshans2003b,jouguet2013,diamanti2007,JosephEXP,ulrik-entropy,CV-MDI-QKD,QIprot2,QIprot3,Madsen},
but also their security analysis has been gradually refined to
incorporate finite-size effects~\cite{Leverrierfsz,rupertPRA,FS
CV-MDI-QKD} and composable
aspects~\cite{leverrierCOMP,leverrierGEN,CMP CV-MDI-QKD}.

We know that Gaussian encoding may be subject to a reduced
performance due to the reconciliation codes. This issue can be
easily fixed by resorting to a discrete-alphabet encoding, e.g.,
coherent states with fixed energy but discrete shifting of their
phase as in Ref.~\cite{phase-shift key}. Nevertheless, the study
of these protocols has been mainly restricted to the case of a
pure-loss channel. In Ref.~\cite{discrete1 -Leverier, discrete2
-Leverier, discrete3 -Leverier} a bound for the secret key rate
has been calculated for two or four coherent states in a thermal
loss channel. However, this was based on a Gaussian
approximation~\cite{GOPT} of the alphabet, which rapidly becomes
loose when the energy of the states increases. Also note that
Refs.~\cite{binary mod,ternary mod} studied binary and ternary
modulation protocols in the presence of collective attacks.

In this work, we consider a multi-letter protocol where the
letters are encoded in different phases of a coherent state with
fixed energy, so as to form a symmetric constellation of coherent
states equidistant from the origin of the phase space. For this
phase-encoded protocol, we compute the secret key rate in direct
and reverse reconciliation assuming a thermal-loss channel, i.e.,
the presence of an entangling cloner collective
attack~\cite{Weedbrook2012,cloner}, which is the most typical and
realistic collective Gaussian attack~\cite{attackG}. We perform an
asymptotic security analysis based on infinitely-many uses of the
channel, so that the secret-key rate may be computed from the
Devetak-Winter formula~\cite{Devetak207}. While our analysis is
for arbitrary $N$ number of phases, we specify the results for the
case of $N=4$ which well approximates the continuous limit
$N\rightarrow\infty$ when the energy of the states is sufficiently
low.

%In Section~\ref{direct reconciliation}, we present the direct
%reconciliation secret key rate of the protocol for the thermal
%loss channel. Before doing so, we calculate an upper bound based
%on the fact that that the parties have a quantum memory and can
%apply an optimal measurement collectively to the signal states
%after the end of the signal exchange. With regard this bound we
%examine the performance of the protocol connected to the
%characteristics of the encoding, i.e., to the modulus of the
%amplitude of the coherent states and the number of them.
%Furthermore, we compare this performance with the performance of a
%protocol with an infinite number of states given by a continuous
%uniform distribution of coherent states on a given radius. We can
%see that for a given range of the radius the protocol with
%discrete encoding and the protocol with the continuous one can
%achieve the same level of performance.
%
%In Section~\ref{reverse reconciliation}, we present the reverse
%reconciliation secret key rate for the thermal loss channel. In
%both cases of the secret key rate calculation, we have included
%the study for the pure loss channel for the sake of completeness.
%%%%%%%%%%%%%%%%%%%%%%%%%%%%%%%%%%%%%%%%%%%%%%%%%%%%%%%%%%%%%%%%%%%%%%%%%%%%%%%%%%%%%%%%%%%

\section{Protocol}
\begin{figure}[pth]
\vspace{-0.2cm} \centering
\includegraphics[width=0.33\textwidth]{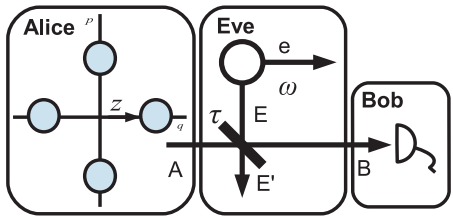}
\caption{\label{fig:scheme}Alice prepares mode A in one of the
four coherent states of the constellation $C(z,4)$ with radius $z$
and sends it to Bob through a thermal-loss channel dilated into an
entangling-cloner attack. In particular, the beam splitter has
transmissivity $\tau$, characterizing the channel loss, and the
variance $\omega\geq 1$ of Eve's TMSV state provides additional
thermal noise to the channel. Eve's output modes are stored in a
quantum memory measured at the end of the protocol, i.e., after
the entire quantum communication and Alice and Bob's classical
communication. At the output of the channel, Bob applies an
heterodyne detection to mode B. An upper bound on the performance
of the parties can be computed by assuming that also Bob has a
quantum memory that he measures at the end of the entire
communication process.}
\end{figure}

Consider a discrete alphabet with $N$ letters, randomly drawn by
Alice. Each letter $k$ is encoded into a coherent state with
amplitude $a_k=z e^{i \phi_k}$, where $z$ is a fixed radius in
phase space (it is just the square root of the mean number of
photons) and the phase is given by $\phi_k=\frac{2\pi}{N}k$. We
call each realization $C(z,N)$ of this encoding scheme a
``constellation''. As an example, a four-state constellation is
shown in Fig.~\ref{fig:scheme}. The coherent state is prepared on
mode $A$ which is sent through a thermal-loss channel, whose
output $B$ is detected by Bob. In a practical realization of the
protocol, this measurement is an heterodyne
detection~\cite{hetero}.

As already mentioned, the thermal-loss channel describes the
effect of an entangling cloner collective attack~\cite{cloner}. In
each use of the channel, Eve's modes $e$ and $E$ are prepared in a
two-mode squeezed vacuum (TMSV) state with variance $\omega\geq
1$, so that
 $\bar{n}=\frac{\omega-1}{2}$ is the mean number of photons in each
 thermal mode~\cite{Weedbrook2012}. Mode $E$ interacts with Alice's mode $A$ via a beam splitter
with transmissivity $\tau$, which characterizes the channel
losses. Eve's output mode $E'$ and kept mode $e$ are then stored
in a quantum memory which is measured at the end of the protocol.
Note that for $\bar{n}=0$ Eve is injecting a vacuum mode, so that
the channel becomes a pure-loss
channel~\cite{Weedbrook2012,PLOB15}. In this case, the output
modes $E'$ and $B$ are described by coherent states with
attenuated amplitudes.
%%%%%%%%%%%%%%%%%%%%%%%%%%%%%%%%%%%%%%%%%%%%%%%%%%%%%%%%%%%%%%%%%%%%%%%%%%%%%%%%%%%%%%%%%%%

\section{Direct reconciliation\label{direct reconciliation}}
We start by presenting the analysis of the protocol in direct
reconciliation~\cite{GG02}, where Bob infers Alice's input. This analysis is
first given for the pure-loss channel, considering an upper bound
for the key rate (assuming a quantum memory for Bob) and then a
realistic key rate (where Bob applies heterodyne detection). We
then generalize the realistic key rate to a thermal-loss channel,
presenting the specific results for $N=4$ coherent states.

\subsection{Pure loss channel}
\subsubsection{Upper bound for the secret key rate}
In this section, we assume that Bob has a quantum memory so that
he may apply an optimal joint detection. This gives an upper bound
to the actual performance of the protocol. This analysis provides
simple results that allow us to give an insight on the performance
with respect to different constellation parameters $z$ and $N$. In
particular, we may show the conditions where $N=4$ coherent
states allow the parties to achieve essentially the same
performance as $N \rightarrow \infty$ coherent states.

\begin{figure}[ht]
\vspace{-0.2cm} \centering
\includegraphics[width=0.45\textwidth]{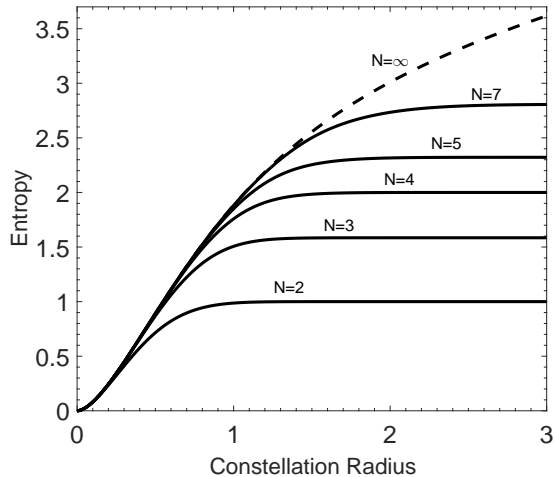}
\caption{\label{fig:entropyfordifferentN} The von Neumann entropy
$S(\rho_A)$ of the Alice's average state $\rho_A$ for different
number $N$ over the radius $z$ of the constellation circle (solid
lines). We plotted also the entropy of the continuous  uniform
distribution ($N\rightarrow\infty$) of the constellation states
(dashed line).}
\end{figure}

Because Alice is sending coherent states $|a_k\rangle$ with the
same probability $p_k=1/N$, the average state before the channel
is given by
\begin{equation}
\rho_A=\frac{1}{N} \sum_{k=0}^{N-1} |a_k\rangle \langle a_k|.
\end{equation}
It is clear that this state is parameterized by $N$ and $z$. In
Fig.~\ref{fig:entropyfordifferentN}, we have plotted the von
Neumann entropy $S(\rho_A)$ of $\rho_A$ for different $N$ over the
radius of the encoding scheme $z$. Recall that
\begin{equation}
S(\rho):=-\text{Tr}(\rho \log_2\rho)=-\sum_j n_j\log_2 n_j,
\end{equation}
where $n_j$ are the eigenvalues of a generic state $\rho$ (see
Appendix~\ref{Orthonormal basis} for more details on how to
compute this entropy via a preliminary Gram-Schmidt procedure).
The entropy $S(\rho_A)$ is larger as we increase the number of
states in the circle. For any given $N$, the entropy saturates to
a constant value after a certain value of the radius $z$. We also
consider the limit of $N\rightarrow\infty$ (see
Appendix~\ref{secLIMIT} for the calculation of the corresponding
average state).

After a pure-loss channel with transmissivity $\tau \in (0,1)$,
Bob's average state will be
\begin{equation}
\rho_B=\frac{1}{N} \sum_{k=0}^{N-1} |\sqrt{\tau} a_k\rangle \langle \sqrt{\tau} a_k|.
\end{equation} Assuming that Bob accesses a quantum memory and may perform a
collective optimal detection of all the output modes, his
accessible information is bounded by the Holevo
information~\cite{Weedbrook2012}
\begin{equation}\label{HolevoAB}
\chi(B:\{a_k\})=S(\rho_B)-\frac{1}{N}\sum_{k=0}^{N-1}S(|a_k\rangle
\langle a_k|).
\end{equation}
In particular, since a coherent state is a pure state its von
Neumann entropy is zero, which simplifies Eq.~\eqref{HolevoAB}
into $\chi(B:\{a_k\})=S(\rho_B)$. In order to calculate the von
Neumann  entropy of the mixture $\rho_B$, we express the $N$
coherent states in terms of a Gram-Schmidt orthonormal basis (see
details in Appendix~\ref{Orthonormal basis}).

In the same fashion, we calculate the Holevo information of the
eavesdropper, who can keep in a quantum memory the other output
$E'$ of the beam splitter. Then Eve's average state will be given
by
\begin{equation}
\rho_{E'}=\frac{1}{N} \sum_{k=0}^{N-1} |\sqrt{1-\tau} a_k\rangle
\langle \sqrt{1-\tau} a_k|
\end{equation}
and her accessible information by
\begin{equation}\label{HolevoE}
\chi(E':\{a_k\})=S(\rho_{E'}).
\end{equation}
Therefore, we get the optimal secret key rate
\begin{equation}\label{upper bound}
R=\chi(B:\{a_k\})-\chi(E':\{a_k\})=S(\rho_B)-S(\rho_{E'}).
\end{equation}

\begin{figure}[t]
\vspace{-0.2cm} \centering
\includegraphics[width=0.45\textwidth]{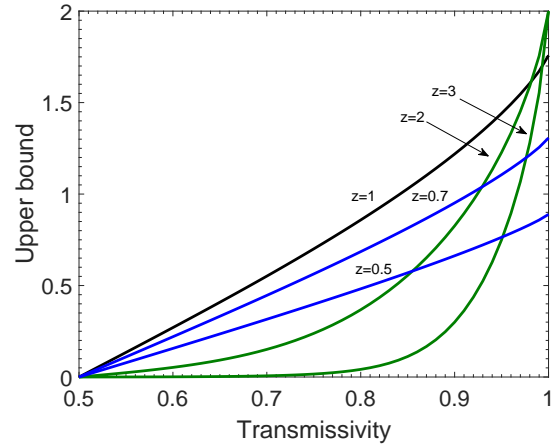}
\caption{The optimal secret-key rate of Eq.~(\ref{upper bound})
for $N=4$ is plotted over the transmissivity $\tau$ for different
values of the radius $z$ of the constellation. We can see that for
values $z<1$ the rate decreases as $z$ is decreasing (blue lines)
while for $z>1$ the rate decreases as $z$ increases till it gets
to zero for $z=10^6$ (green lines).}\label{figg}
\end{figure}

\begin{figure}[t]
\vspace{-0.2cm} \centering
\includegraphics[width=0.45\textwidth]{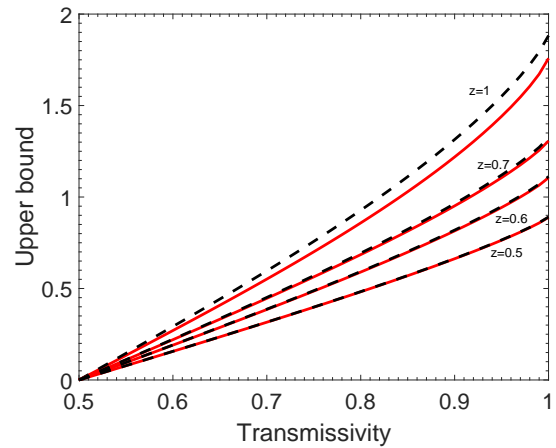}
\caption{ The optimal secret-key rate of Eq.~(\ref{upper bound})
for $N=4$ is plotted over the transmissivity $\tau$ for different
values of the radius $z$ (solid red lines). Here we also plot the
optimal secret key rate for the continuous uniform distribution of
states (black dashed lines). We see that  for $z<0.6$, the two
rates become almost identical. This corresponds to a saturation
point for the 4-state protocol, so that it makes no difference to
use four coherent states or infinite.}\label{figg2}
\end{figure}

In Fig.~\ref{figg} we plotted this optimal rate for $N=4$ as a
function of the transmissivity $\tau$ and for different values of
the radius $z$. We see that there is an optimal intermediate value
for $z$, so that it cannot be too small (so that all the coherent
states are too similar to the vacuum), neither too large (so that
all the coherent states become almost-perfectly distinguishable).
Then, in Fig.~\ref{figg2}, we also show that the optimal
performance for the $N=4$ protocol is very close to that of the
continuous-alphabet protocol $N=\infty$ for the relevant values of
the radius $z$.

%Subsequently, the average states $\rho_B$ and $\rho_E$ are dependent on $z$ and $N$ as the output of the pure loss with input the average state $\rho_A$. However, the amplitude of the states are dependent also on the transmissivity of the channel $\tau$. This means that for a particular $N$ if the entropies in Eq.~\eqref{upper bound} have surpassed the saturation point will be almost equal resulting in a vanishing secret key rate.
%For instance, for $N=4$, we observe that for approximately $z>1.5$ the entropy is almost constant. In turn, this means that the entropies for $z^\prime=\sqrt{\tau}z$ and $z^{\prime \prime}=\sqrt{1-\tau}z$ will be approximately equal when $\tau<0.75$ and $z=3>1.5$. We can observed this in Fig.~\ref{fig:rateG=4differentz}, where for $N=4$ and $z=3$ the upper bound for transmissivities less than $\tau=0.75$ vanishes.

%Eve's entropy is increasing as she can distinguish easier the signal states of Alice. More specifically, as their distance in their phase space and consequently their overlap is as small as possible. The goal is to make sure that even for large values of $\bar{\tau}=1-\tau$, low transmissivities, this is possible.

%As the radius of the constellation is becoming larger their
%overlap is decreasing. However, if we increase the number of
%states in a given radius then the overlaps between each other are
%increasing as well. Therefore, a critical point is to investigate
%for which radii a protocol with a given number of states is
%possible to achieve a performance very close to the case where the
%number of states is asymptotically going to infinity. In
%particular, in Fig.~\ref{fig:entropyfordifferentN}, we can see
%that for a specific range of $z$ the entropy for $N$ states
%coincides with the entropy of the state given by a uniform
%distribution of coherent states as in Eq.~\eqref{uniform
%distribution}. As a consequence, there is a range of $z$ that the
%bound of the secret key rate for a given $N$ almost coincides with
%that of the continuous uniform distribution of states. This is
%illustrated in Fig.~\ref{fig:smaleestradius}, where we have
%plotted the bounds for the case of $N=4$ (red solid lines) and for
%$N\rightarrow\infty$ (black dashed lines) for different values of
%$z$.
%%%%%%%%%%%%%%%%%%%%%%%%%%%%%%%%%%%%%%%%%%%%%%%%%%%%%%%%%%%%%%%%%%%%%%%%%%%%%%%%%%%%%%%%%%%

\subsubsection{Realistic secret key rate \label{DR lower-bound}}
Contrary to the previous discussion, the realistic situation is
dictated by the limitations in the current technology. In this
case, Bob does not use a quantum memory and an optimal collective
measurement but individual heterodyne detections, with a
continuous (complex) outcome $b$. Therefore, in order to calculate
the secret key rate, we need to consider the corresponding mutual
information between Alice and Bob. Let us define the variables
$X_A=\{a_k, p_k\}$ with $p_k=1/N$ and $X_B=\{b, p(b)\}$. Then, we
consider
\begin{equation}\label{mutual info}
I(X_A:X_B)=H(X_A)-H(X_A|X_B),
\end{equation}
where $H$ is the Shannon entropy and $H(...|...)$ the conditional
Shannon entropy. Recall that
\begin{equation}
H(X_A|X_B)=\int p(b) H(X_A|X_B=b)d^2b.
\end{equation}

\begin{figure}[t]
\vspace{-0.2cm} \centering
\includegraphics[width=0.45\textwidth]{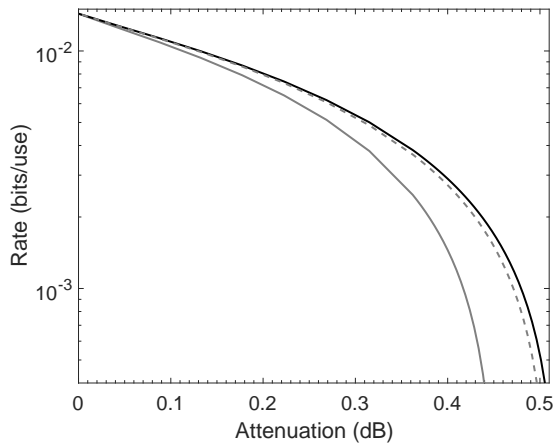}
\caption{Realistic secret-key rate (bits/use) over the attenuation
(decibels) in direct reconciliation for $N=4$ and $z=0.1$. We plot
the rate for a pure-loss channel (upper solid line) and a
thermal-loss channel with mean photon number $\bar{n}=0.01$
(middle dashed line) and $\bar{n}=0.1$ (lower solid
line).}\label{figDR1}
\end{figure}

It is clear that $H(X_A)=\text{log}_2N$. In order to calculate the
probability distribution $p(a_k|b)$, i.e., the probability that
the state $|a_k\rangle$ was sent through the channel given that
Bob measured the amplitude $b$. The probability that Bob measures
$b$ given that the coherent state $|\alpha_k\rangle$ was sent
through the channel is given by $p(b|a_k)=\frac{1}{\pi}
e^{-|b-\sqrt{\tau}a_k|^2}$. Therefore, we can apply Bayes' rule to
obtain
\begin{equation}\label{PKB}
p(a_k|b)=\frac{1}{N \pi p(b)}e^{-|b-\sqrt{\tau}a_k|^2},
\end{equation}
where $p(b)=\frac{1}{N}\sum_{k=0}^N p(b|a_k)$. With all these
elements we can compute the Devetak-Winter rate
$R=I(X_A:X_B)-S(\rho_{E'})$ which is plotted in Fig.~\ref{figDR1}
for $N=4$.

%%%%%%%%%%%%%%%%%%%%%%%%%%%%%%%%%%%%%%%%%%%%%%%%%%%%%%%%%%%%%%%%%%%%%%%%%%%%%%%%%%%%%%%%%%%

\subsection{Thermal loss channel \label{DR lower-bound thermal}}
We now consider the more general case of a thermal-loss channel,
i.e., the presence of an entangling-cloner attack. Let us write
Eve's TMSV state in the Fock basis~\cite{Weedbrook2012}
\begin{equation}
\rho_{Ee}(\lambda)=(1-\lambda^2)\sum_{n=0}^\infty
(-\lambda)^{(k+l)} |k\rangle\langle l|\otimes|k\rangle\langle l|,
\end{equation}
with
$\lambda=\text{tanh}\left[\frac{1}{2}\text{arcosh}(2\bar{n}+1)\right]$,
where $\bar{n}$ is the mean number of thermal photons. Let us
apply the beam splitter operation to Alice's mode $A$ and Eve's
mode $E$, with annihilation operators $\hat{a}_A$ and $\hat{a}_E$,
respectively. This is given by~\cite{Weedbrook2012}
\begin{equation}
U(\theta)=\text{exp}\left[\theta
\left(\hat{a}_A^\dagger\hat{a}_E-\hat{a}_A\hat{a}_E^\dagger
\right)\right],
\end{equation}
where $\theta=\text{arcos}(\sqrt{\tau})$. Therefore, the global
output state of Bob (mode $B$) and Eve (modes $e$ and $E'$), is
given by
\begin{equation}
\rho_{BE'e}(\theta,a_k,\lambda)=U(\theta)\Pi_\text{A}(a_k)\rho_{Ee}(\lambda)U^\dagger(\theta),
\end{equation}
where  $\Pi_\text{A}(a_k):= |a_k\rangle \langle a_k|$. By tracing
out $B$, we obtain Eve's state
\begin{equation}\label{conditional Eve}
\rho_{\text{Eve}|k}:=\rho_{\text{E'e}}(\theta,a_k,\lambda)=\text{Tr}_\text{B}[\rho_{BE'e}(\theta,a_k,\lambda)].
\end{equation}

The average state of Eve is given by the convex sum
\begin{equation}\label{Eve's average state}
\rho_{\text{Eve}}(\theta,z,\lambda)=\frac{1}{N}\sum_{k=0}^{N}\rho_{\text{Eve}|k}.
\end{equation}
Therefore, the Holevo information is given by
\begin{equation}\label{HolevoE1}
\chi(\text{Eve}:X_A)=S(\rho_{\text{Eve}})-\frac{1}{N}\sum_{k=0}^N
S(\rho_{\text{Eve}|k}).
\end{equation}
The entropy of the state $\rho_{\text{Eve}|k}$ does not depend on
$k$, i.e., the phase of the amplitude of the coherent state that
Alice has sent. Thus Eq.~\eqref{HolevoE1} can be simplified to
\begin{equation}\label{HolevoE2}
\chi(\text{Eve}:X_A)=S(\rho_{\text{Eve}})-S(\rho_{\text{Eve}|k}),
\end{equation}
for any $k$. In order to calculate the mutual information, we
follow the reasoning of Section~\ref{DR lower-bound} with the
difference that Bob's probability distribution is given by
\begin{equation}\label{PBKth}
p(b|a_k)(\bar{n})=\text{Tr}[\Pi(b)\rho(\sqrt{\tau}a_k,(1-\tau)\bar{n})\Pi^\dagger(b)],
\end{equation}
where $\Pi(b):= |b\rangle \langle b|$ and
$\rho(\sqrt{\tau}a_k,(1-\tau)\bar{n})$ is a displaced thermal
state with amplitude $\sqrt{\tau}a_k$ and mean photon number
$(1-\tau)\bar{n}$. We find (see Appendix~\ref{displaced thermal
state})

\begin{equation}\label{PKBth}
p(b|a_k)(\bar{n})=\frac{\text{exp}\left[\frac{|b-\sqrt{\tau}a_k|^2}{1+(1-t)\bar{n}}\right]}{\pi
(1+(1-\tau)\bar{n})}.
\end{equation}
Using the Bayes' rule we can derive $p(a_k|b)(\bar{n})$ and
compute Alice and Bob's mutual information via the formula in
Eq.~(\ref{mutual info}). Altogether, we then compute (numerically)
the direct reconciliation secret-key rate
\begin{equation}\label{lower thermal}
R(\bar{n})=I(X_A:X_B)(\bar{n})-\chi(\text{Eve}:X_A).
\end{equation}

In Fig.~\ref{figDR1}, we plot this secret key rate over the
attenuation for a protocol with $N=4$ and $z=0.1$. In particular,
we see that the performance obtained in the presence of thermal
noise $\bar{n}=0.01$ is not so far from the performance achievable
in the presence of a pure-loss channel. In other words, the
four-state protocol is sufficiently robust to the presence of excess
noise. However, as expected, we also have that direct
reconciliation restricts the use of the protocol to low loss.
The case is for different reverse reconciliation that we study
below.

%%%%%%%%%%%%%%%%%%%%%%%%%%%%%%%%%%%%%%%%%%%%%%%%%%%%%%%%%%%%%%%%%%%%%%

\section{Reverse reconciliation \label{reverse reconciliation}}
As before, for the sake of simplicity, we start by considering the
case of a pure-loss channel in reverse reconciliation~\cite{Grosshans2003b}
and then we extend the results to the presence of thermal noise.
We just need to re-compute Eve's Holevo
bound (now with respect to Bob's outcomes). More specifically, we
need to re-compute Eve's conditional entropy.

\begin{figure}[t]
\vspace{-0.2cm} \centering
\includegraphics[width=0.45\textwidth]{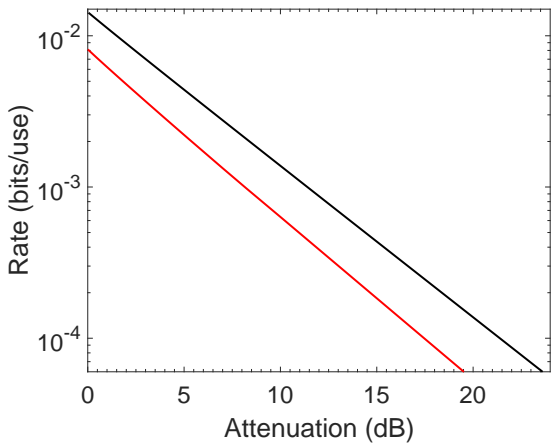}
\caption{Realistic secret key rate (bits/use) over the attenuation
(decibels) in reverse reconciliation for $N=4$ and $z=0.1$. We
have plotted the rate for a pure-loss channel (upper black line)
and a thermal-loss channel with excess noise $\epsilon=0.001$
(lower red line). Both these rates coincide with the corresponding
rates achievable by a Gaussian protocol modulating coherent states
with variance $V_M=0.02$.}\label{RRFig1}
\end{figure}

Eve's state conditioned to Bob's outcome $b$ is
\begin{equation}
\rho_{E'|b}=\sum_{k=0}^{N-1}
p(a_k|b)|\sqrt{1-\tau}a_k\rangle\langle\sqrt{1-\tau}a_k|,
\end{equation}
where $p(a_k|b)$ is given in Eq.~\eqref{PKB}. We can then compute
$S(\rho_{E'|b})$ which is now depending on $b$. Using this
quantity, we may write the secret-key rate
\begin{equation}\label{RR rate}
R=I(X_A:X_B)-S(\rho_{E'})+\int d^2b p(b) S(\rho_{E'|b}).
\end{equation}
This rate is plotted in Fig.~\ref{RRFig1} for the four-state
protocol $N=4$ and radius $z=0.1$.
%where we have replaced in Eq.~\eqref{DEVETAKCRITERION}
% $I_\text{AB}$ with $I(A:B)$ from from Eq.~\eqref{mutual info} and $I_E$ with $\chi(E:\beta)$ from Eq.~\eqref{HolevoEB}.
%%%%%%%%%%%%%%%%%%%%%%%%%%%%%%%%%%%%%%%%%%%%%%%%%%%%%%%%%%%%%%%%%%%%%%%%%%%%%%%%%%%%%%%%%%%%%%%%%%%%%%%%%%%%%%
%Here we continue from the discussion of Section~\ref{DR lower-bound thermal}. Eve's average state $\rho _{\text{Eve}}$ is given in Eq.~\eqref{Eve's average state}.

Let us now consider the presence of thermal noise. In this case,
Eve's conditional state is given by
\begin{equation}
\rho_{E'e|b}=\sum_{k=0}^{N-1}p(a_k|b)(\bar{n})\rho_{\text{Eve}|k},
\end{equation}
where $\rho_{\text{Eve}|k}$ is given in Eq.~\eqref{conditional
Eve} and $p(a_k|b)$ comes from Eq.~(\ref{PKBth}). Therefore, we
may derive $S(\rho_{E'e|b})$ and calculate the secret-key rate
\begin{equation}\label{RR rate thermal}
R(\bar{n})=I(X_A:X_B)(\bar{n})-S(\rho _{\text{Eve}})+\int d^2b~
p(b)(\bar{n})\rho_{E'e|b},
\end{equation}
where
$p(b)(\bar{n}):=\frac{1}{N}\sum_{k=0}^{N-1}p(b|a_k)(\bar{n})$.
Numerically, we compute this rate by truncating the Hilbert space
to a suitable number of photons, which is of the order of $\simeq
10-15$ photons for the specific regime of parameters considered.

\begin{figure}[ht]
\vspace{-0.1cm} \centering
\includegraphics[width=0.45\textwidth]{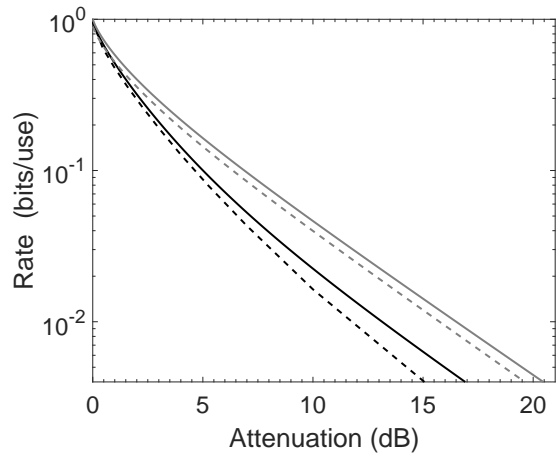}
\caption{Realistic secret key rate (bits/use) over the attenuation
(decibels) in reverse reconciliation over the attenuation
(decibels) for $N=4$ and $z=1$. We have plotted the rate for a
pure-loss channel (lower solid line) and a thermal-loss channel
with excess noise $\epsilon=0.01$ (lower dashed line). The
corresponding secret key rate for the protocol with Gaussian
modulation ($V_M=2$) has also been plotted for the case of
pure-loss channel (upper solid line) and thermal-loss channel with
excess noise $\epsilon=0.01$ (upper dashed line). We see that, for
this regime of energies, the rate of the four-state protocol does
not coincide with the rate of the Gaussian
protocol.}\label{RRFig2}
\end{figure}

In Fig.~\ref{RRFig1}, we plot the reverse reconciliation secret
key rate over the attenuation for the four-state protocol $N=4$
with radius $z=0.1$ and excess noise $\epsilon=0.001$~\cite{note}.
We can see that the protocol is sufficiently robust to excess
noise, achieving a rate of $6\times 10^{-4}$ bits per channel use
for attenuation values of about $20$~dB. In this regime of energy,
the performance of the protocol coincides with that of a Gaussian
protocol modulating coherent state with modulation variance $V_M=2
z^2$ (and performing heterodyne detection on the channel output).
On the contrary, for larger energies, e.g., for a constellation
radius $z=1$, the rate of the four-state protocol does not
coincide with its Gaussian counterpart, as also illustrated in
Fig.~\ref{RRFig2}. Here the four-state protocol can achieve a rate
of the order of $4\times 10^{-3}$ bits per channel use for
attenuation values of about $15$~dB and excess noise
$\epsilon=0.01$.
%%%%%%%%%%%%%%%%%%%%%%%%%%%%%%%%%%%%%%%%%%%%%%%%%%%%%%%%%%%%%%

\section{Conclusion}
In this work, we have investigated finite-alphabet coherent-state
QKD protocols, where the encoding is performed by randomly
choosing the phase of the coherent states so that they are
iso-energetic and symmetrically distributed around the origin of
the phase space. Considering an optimal scenario where Bob may
access a quantum memory and the channel is pure-loss, we have
analyzed the conditions under which the use of four states can
approximate a continuous alphabet. Our analysis is asymptotic,
i.e., we assume the limit of infinite signal states exchanged by
the remote parties, so that it does not account for finite-size
effects and composable aspects. Nevertheless, this is the first
study of these types of protocols in the presence of realistic
thermal-loss conditions, without assuming Gaussian approximations.
In reverse reconciliation, we find that the four-state
phase-encoded protocol is sufficiently robust to loss and noise,
so that it may be used to extract secret keys at metropolitan
mid-range distances (e.g. around $75$~km).
%%%%%%%%%%%%%%%%%%%%%%%%%%%%%%%%%%%%%%%%%%%%%%%%%%%%%%%%%%%%%%%%%%%%%%

\section{Acknowledgements}
C.W. would like to acknowledge the Office of Naval Research
program Communications and Networking with Quantum
Operationally-Secure Technology for Maritime Deployment
(CONQUEST), awarded to Raytheon BBN Technologies under prime
contract number N00014-16-C-2069. P. P. acknowledges support from
the EPSRC via the `UK Quantum Communications Hub' (EP/M013472/1)
and would like to thank Thomas Cope for advices on the use of the
computer cluster of the University of York (YARCC). C. L.
acknowledges support from Innovation Fund Denmark (Qubiz project).

%The content of this paper does not necessarily reflect the
%position or policy of the Government and no official endorsement
%should be inferred

%%%%%%%%%%%%%%%%%%%%%%%%%%%%%%%%%%%%%%%%%%%%%%%%%%%%%%%%%%%%%%%%%%%%%%%%%%%%%%%%%%%%%%%%%%%%%%%%%%%%%%%%%%%%

\appendix

\section{Orthonormal basis for $N$ coherent states \label{Orthonormal basis}}
Suppose that we have  $N$ coherent
states described by amplitudes $a_k$ for $k=0,1\dots N-1$. Since
these states are non-orthogonal we can have a matrix $\mathbf{V}$
that describes their overlaps, which are given by
\begin{equation}
V_{ij}=\langle a_i|a_j\rangle=\text{exp}\left
[-\frac{1}{2}\left(|a_i|^2+|a_j|^2-2a_i^*a_j\right)\right ].
\end{equation}
For a constellation of states as described before and after the
attenuation due to the propagation through a pure-loss channel,
the overlaps for Bob are given by
\begin{equation}
V^B_{ij}=\langle \sqrt{\tau}a_i| \sqrt{\tau}
a_j\rangle=\text{exp}\left [\tau z^2 \left(e^{{\textrm{i}} \frac{2
\pi}{N}(j-i)}-1\right)\right],
\end{equation}
while for Eve we may write
\begin{align}
V^E_{ij}=&\langle \sqrt{1-\tau}a_i| \sqrt{1-\tau}
a_j\rangle=\nonumber\\&=\text{exp}\left[(1-\tau)z^2
\left(e^{\textrm{i}
 \frac{2 \pi}{N}(j-i)}-1\right)\right].
\end{align}
Then, according to the Gram-Schmidt procedure, we can derive an
orthonormal basis $\{
|i\rangle\}=\{|0\rangle,|1\rangle,\dots|N-1\rangle\}$ for the
subspace spanned by these $N$ coherent states. As a result, each
state will be expressed as a superposition of this basis vectors
as
\begin{equation}
|a_k\rangle=\sum_{i=0}^k M_{ki} |i\rangle
\end{equation}
where the $M_{ki}$ can be computed by the algorithm
\begin{itemize}
\item[] $M_{k0}=V_{0k}$,
\item[] $M_{ki}=\frac{1}{M_{ii}}\left(V_{ik}-\sum_{j=0}^{i-1} M^*_{ij}M_{kj}\right)$ if $1 \leq i<k$,
\item[] $M_{ki}=0$ otherwise,
\item[] $M_{kk}=\sqrt{1-\sum_{i=0}^{k-1} |M_{ki}|^2}$ for $k>0$.
\end{itemize}
Then the density matrix $\rho(a_k)=|a_k\rangle\langle a_k|$ is
given by
\begin{equation}
\rho(a_k)=\sum_{i,j=0}^k M_{k,i}M^*_{k,j} |i\rangle \langle j |,
\end{equation}
and the average state takes the form
\begin{equation}
\rho=\frac{1}{N}\sum_{k=0}^{N-1}\rho(a_k)=\frac{1}{N}\sum_{k=0}^{N-1}\sum_{i,j=0}^k
M_{k,i}M^*_{k,j} |i\rangle \langle j |.
\end{equation}
Diagonalizing the previous state, we then compute its von Neumann
entropy.
%%%%%%%%%%%%%%%%%%%%%%%%%%%%%%%%%%%%%%%%%%%%%%%%%%%%%%%%%

\section{Asymptotic state for a continuous alphabet}\label{secLIMIT} Let us express
a coherent state in the Fock basis, i.e,
\begin{equation}\label{Fock Basis}
\Pi(a):=|a \rangle\langle a|=e^{-|a|^2}\sum_{n,m=0}^{\infty}
\frac{a^n (a^\dagger)^m}{\sqrt{n!}\sqrt{m!}}|n \rangle\langle m|
\end{equation}
In order to be able to do numerical calculations, we have to
truncate the Fock space and a very good approximation is given by
$n\sim 2 |\alpha|^2$. As a result, in this truncated Fock basis,
the state will be
\begin{equation}
\Pi^{\text{tranc}}(a)\simeq e^{-|a|^2}\sum_{n,m=0}^{2
\left\lfloor |a|^2 \right\rfloor} \frac{a^n
(a^\dagger)^m}{\sqrt{n!}\sqrt{m!}}|n \rangle\langle m|.
\end{equation}

For $N$ coherent states in a constellation with radius $z$, the
average state can be written as
\begin{equation}\label{constelationinfock}
\rho=\frac{e^{-z^2}}{N}\sum_{n,m=0}^{2  \left\lfloor z^2
\right\rfloor}  \frac{z^{(n+m)}\sum_{j=0}^{N-1} e^{\textrm{i}
\frac{2 \pi}{N} (n-m)j}}{\sqrt{n!}\sqrt{m!}}|n \rangle\langle m|,
\end{equation}
where the non zero terms are the terms with $m-n=N$ and $n=m$. For
a continuous distribution $p(a_\phi)=\frac{1}{2 \pi}$ of
phase-encoded coherent states $|a_\phi \rangle$ with fixed radius
$z=|a|$ and $\phi=\text{arg}(a_\phi)$,
Eq.~(\ref{constelationinfock}) becomes
\begin{align}\label{uniform distribution}
\rho=&\frac{e^{-z^2}}{2 \pi}\sum_{n,m=0}^{2  \left\lfloor z^2 \right\rfloor}  \frac{z^{(n+m)}\int_{0}^{2 \pi}   e^{\textrm{i} \phi (n-m)} d\phi}{\sqrt{n!}\sqrt{m!}}|n \rangle\langle m|=\nonumber\\
=&e^{-z^2}\sum_{n=0}^{2  \left\lfloor z^2 \right\rfloor}
\frac{z^{2n}}{n!}|n \rangle\langle n|.
\end{align}
%%%%%%%%%%%%%%%%%%%%%%%%%%%%%%%%%%%%%%%%%%%%%%%%%%%%%%%%%%%%%%%%%%%%%%%%%%%%%%%%%%

\section{Displaced thermal state \label{displaced thermal state}}
A thermal state with mean number of photons $\bar{n}$ may be
expressed as a convex sum of coherent states $|a\rangle$ according
to the P-Glauber representation as
\begin{equation}
\rho(\bar{n})=\int p(a,\bar{n})|a\rangle\langle a|d^2a,~~
p(a,\bar{n})=\frac{1}{\bar{n} \pi} e^{-|a|^2/\bar{n}}.
\end{equation}
Applying the displacement operator $D(d)$, which displaces a
coherent state $|a\rangle$ with amplitude $a$ into a coherent
state $|a+d\rangle$ with amplitude $a+d$, we obtain a displaced
thermal state
\begin{align}
\rho(d,\bar{n})&=D(d)\rho(\bar{n})D^\dagger(d)=\nonumber\\
&=\int p(a,\bar{n})D(d)|a \rangle\langle a|D^\dagger(d)~d^2 a=\nonumber\\
&=\int p(a,\bar{n})|a+d\rangle\langle a+d|d^2 a=\nonumber\\
&=\int p(c-d,\bar{n})|c\rangle\langle c|d^2 c
%&=\int p(\beta,\bar{n})|\gamma\rangle\langle\gamma|d^2\gamma
\end{align}
with $p(c-d,\bar{n})=\frac{1}{\bar{n} \pi} e^{-|c-d|^2/\bar{n}}.$
According to equation Eq.~(\ref{Fock Basis}), we can have a
representation of this state in Fock basis, so that
\begin{equation}
\rho(d,\bar{n})=\int \sum_{n,m=0}^{\infty}p(a-d,\bar{n})
e^{-|a|^2} \frac{a^n (a^*)^m}{\sqrt{n!}\sqrt{m!}}|n \rangle\langle
m|~d^2 a
\end{equation}
The state after projecting to a coherent state $|b\rangle$
(heterodyne measurement), i.e.,
$\Pi(b)\rho(d,\bar{n})\Pi^\dagger(b)$, will be calculated as
\begin{align}
&\int d^2 a~\sum_{n,m,k,l,i,j=0}^{\infty} p(a-d,\bar{n}) e^{-|\alpha|^2} \frac{\alpha^n (\alpha^*)^m}{\sqrt{n!}\sqrt{m!}}e^{-|b|^2} \frac{b^k (b^*)^l}{\sqrt{k!}\sqrt{l!}}\times \nonumber\\
&\times e^{-|b|^2} \frac{b^i (b^*)^j}{\sqrt{i!}\sqrt{j!}}|k \rangle\langle l||n \rangle\langle m||i \rangle\langle j|=\\
&\int d^2 a~p(a-d,\bar{n}) e^{-|a|^2}e^{-2|b|^2}\sum_{n,m=0}^{\infty} \frac{(a b^*)^n (b a^*)^m}{\sqrt{n!}\sqrt{m!}}\times\nonumber\\
&\times \sum_{k,j=0}^{\infty}\frac{b^k
(b^*)^j}{\sqrt{k!}\sqrt{j!}}|k \rangle\langle j|,
\end{align}
and, applying the trace operation, we obtain the probability
distribution
\begin{align}
p(b|d)(\bar{n})&=\int d^2a~\frac{1}{\bar{n} \pi} e^{-|a-d|^2/\bar{n}}e^{-(|a|^2+|b|^2-b^*a-b a^*)}=\nonumber\\
&=\frac{1}{\bar{n} \pi}\int e^{-|a-d|^2/\bar{n}} e^{-|a-b|^2} d^2a=\nonumber\\
&=\frac{1}{(\bar{n}+1)\pi}\text{exp}\left(-|b-d|^2/(\bar{n}+1)\right).
\end{align}
Let us write this probability distribution for the thermal output
state of a thermal-loss channel with transmissivity $\tau$ and
mean thermal photon number $\bar{n}$ when applied to an input
coherent state $|a_k\rangle$ ($d:=\sqrt{\tau}a_k$). We find
Eq.~(\ref{PKBth}).


\begin{thebibliography}{99}
\bibitem{note} In order to evaluate the excess noise of the
thermal-loss channel, Alice may send Gaussian-modulated decoy
coherent states to Bob (interleaved with coherent states used for
the key). At the end of all quantum communication, Alice informs
Bob in which instances she was sending decoys so that they can use
their data to compute $\epsilon$. In the asymptotic regime, we may
write $\epsilon=\frac{(1-\tau)(\omega-1)}{\tau}$.
\end{thebibliography}
\end{document}